\documentstyle[multicol,aps,graphicx]{revtex}
\title{Diffusion-reaction in thermal 
growth of silicon oxide films on Si}
\author{R.M.C. de Almeida$^*$, S.Gon\c{c}alves$^*$, I.J.R. Baumvol$^*$, 
and F.C. Stedile$^{**}$\\ $^*$Instituto de F\'{\i}sica and $^{**}$Instituto de
 Qu\'{\i}mica\\
 Universidade Federal do Rio Grande do Sul \\ 
Av. Bento Gon\c{c}alves 9500, Porto Alegre, RS  91509-900 \\ 
Brazil}

\begin{document}
\maketitle

\begin{abstract}
The thermal growth of silicon oxide films on Si in dry O$_2$ is modelled as a dynamical system, assuming that it is basically a diffusion-reaction phenomenon. Relevant findings of the last decade are incorporated, as structure and composition of the oxide/Si interface and O$_2$ transport and reaction at initial stages of  growth. The present model departs from the well established Deal and Grove framework (Deal, B.E. and Grove, A. S. General Relationship for the Thermal Oxidation of Silicon {\it J. Appl. Phys.} {\bf 36},  3770-3778  (1965)) indicating that its basic assumptions, steady-state regime and  reaction between O$_2$ and Si at a sharp oxide/Si interface are only attained asymptotically. Experimental growth kinetics by various authors, obtained for a wide range of growth parameters are shown to collapse into one single curve when the scaling properties of this model equations are explored.
\end{abstract}

\begin{multicols}{2}
Silicon oxide films thermally grown on single-crystalline Si (c-Si)
wafers in dry O$_2$ are still the most common materials used as gate
dielectrics in Si-based metal-oxide-semiconductor (MOS) 
structures. As device dimensions shrink below  0.25 $\mu$m, 
the oxide films are forced to scale down to thicknesses of 5 nm 
and less. Reliability of  highly integrated  Si devices is 
critically dependent on the characteristics of the vitreous oxide film, 
like thickness uniformity, defect density, dielectric strength, 
and others, as well as on those of the oxide/Si interface, 
like roughness, electronic states density and others \cite{buch,feld}.

The dielectric performance of ultrathin films of silicon 
oxide thermally grown on Si and the structural and 
electronic properties of the
oxide/Si interface  have been 
intensively studied experimentally, theoretically, and computationally \cite{schu,pasq} in  
synergism with the development of the semiconductor industry. 
The understanding of film growth kinetics, however, did not make any significant 
progress since the early model of Deal and Grove \cite{deal}, which 
subsisted  as the only fundamental approach. The Deal and 
Grove model, and more specifically the resulting linear-parabolic law,  
agrees with the observed growth kinetics in dry O$_2$ only above a 
certain thickness \cite{mass,han,rigo}. This is an expected fact from the 
model itself, since its basic assumption is that growth is promoted 
by interstitial diffusion of the oxidant species, the O$_2$ molecule, 
through a previously grown oxide layer, thick enough to guarantee 
steady state regime. Besides, reaction between O$_2$ and Si is assumed to take 
place solely at an abrupt oxide/Si interface, producing stoichiometric SiO$_2$. This initial oxide thickness 
was estimated \cite{mass,deal} from experimental data as being between 
20 and 30 nm, well above the 
thickness range of interest for present and future use in microelectronics 
Si-devices. 
Description of the
 growth kinetics in the lower thickness range  was addressed by many 
authors, within the Deal and Grove model framework, adding new terms to 
the linear-parabolic expression in order to fit experimental data \cite{mass,han}. Although 
fitting the experimental curves, and so providing useful 
analytical expressions capable of reproducing the whole thickness interval,  
the extra terms added to the Deal and Grove expression did not have a 
well-defined physical meaning, even though their dependence on  some processing 
parameters, like temperature for instance, could be explored \cite{mass}. None of these models has been shown to be 
clearly correct 
and none of them has gained widespread acceptance. One is left then 
with several empirical expressions which can be used to model the growth 
kinetics in dry 
O$_2$ in the thin and ultrathin film regimes below 20 nm, as well as in further stages of 
growth, but with no satisfactory physical explanation \cite{plum}. 
According to B. Deal \cite{deal2} ``such an expression may well fit the data, 
but provides little guidance or insight into mechanisms involved or effects of process variations''.

Recent investigations showed that as film 
thickness decreases below 20 nm, 
the contribution to film growth  due to reaction away from the interface region
becomes increasingly significant  \cite{han2,trim,roch,trim2,lu,hama}. Furthermore, there are strong theoretical and experimental evidences of the existence of a reactive layer formed by sub-oxides (also called Si-excess) near the oxide/Si interface \cite{mott,ston,himp,feld2}.

In  the present work   the thermal growth of silicon oxide films on 
Si is modelled as a dynamical system, assuming that it is basically a diffusion-reaction phenomenon. 
Since steady-state regime is not imposed, an  initial oxide thickness is not required.  Therefore,
 the model is expected 
to describe the whole oxide thickness interval. Diffusion-reaction equations have been used to describe different systems \cite{matt,larr,bark} but, to our knowledge,
this is the first time they are applied to silicon  oxide growth 
kinetics. The diffusing species
 is taken to be O$_2$,  to model what has been largely demonstrated 
by isotopic substitution experiments \cite{han2,trim,pret,guse,guse2}. Growth is promoted by reaction of O$_2$ with Si, not necessarily at the oxide/Si interface, but wherever the two species meet. 
The growth  kinetics can then be obtained at any temperature by 
specifying the diffusivity of the O$_2$ molecule in the silica network, 
$D$, the reaction rate between O$_2$ and Si, $k$, and the O$_2$ 
pressure in the gas phase. Growth in one dimension is considered,  which is the most common experimental situation,
meaning that as a 
face of c-Si is exposed to O$_2$, 
the silica film grows in the direction perpendicular to 
this face. The proposed description for the growth kinetics 
is contained in the following coupled partial differential
 equations:
\begin{eqnarray}
\label{edif}
\frac{\partial{\rho_{O_2}}}{\partial{t}} &=& D  \frac{\partial^2{\rho_{O_2}}}{\partial x ^2} - k \rho_{O_2} \rho_{Si}
 \nonumber \\
\frac{\partial{\rho_{Si}}}{\partial{t}} &=&  -  k \rho_{O_2} \rho_{Si}     
\end{eqnarray}
where $\rho_i = c_i / c_{Si}^{bulk}$  is the concentration of 
the i-species ($i=Si$, $O_2$) in the solid phase, 
in units of number of atoms per unit volume, $c_i$, normalised 
by the concentration of Si in c-Si, $c_{Si}^{bulk}$. Thus, the state 
of any species at any time is characterised by a density function 
$\rho(x,t)$, where $x$ is the coordinate in the 
direction of growth, $x=0$ indicating the surface of the sample, and $t$ is 
the elapsed growth time. 
The first equation describes the O$_2$ concentration 
rate and it is essentially a balance equation. At a given position 
the concentration of oxygen may vary due to the net flux of O$_2$
through fully (SiO$_2$ in oxide bulk) or partially (SiO$_x$, $x < 2$,  near the interface 
\cite{pasq,hama,himp,feld2})
oxidised Si, modelled by the 
diffusion  term  with a 
diffusion constant $D$, or due to reaction 
to produce silicon oxide, characterised  by a reaction rate constant 
$k$. The second equation describes the Si 
concentration rate, which is only due to reaction with O$_2$, implying that 
Si is an immobile species \cite{pret}. The silicon oxide concentration, 
$\rho_{oxide}$,
does not explicitly appear in these equations, but it
is in direct connection 
with $\rho_{Si}$ through the expression 
\begin{equation}
\label{oxi}
\rho_{oxide}(x,t) = 1 - \rho_{Si}(x,t)
\end{equation}
reflecting conservation of Si species in the 
Si + O$_2 \rightarrow$ SiO$_2$ reaction.
It is assumed that the O$_2$ diffusivity at a certain temperature 
does not change during oxidation, and therefore $D$ is constant throughout 
the whole growth 
process. Furthermore, based on the fact that oxygen  diffuses mainly through 
oxide, one neglects volume changes due to the above chemical
reaction. 
In order to complete the description in terms of equations 
(\ref{edif}), it is necessary to provide the initial and boundary 
conditions. They are
\begin{eqnarray}
\label{icon}
\rho_{Si}(x,0) &=& 1       \;\;\;\;\;      \forall x \geq 0  \\
\rho_{O_2} (0,t) &=& \frac{c_{gas}f_v}{c_{Si}^{bulk}}  \;\;\;\;\    \forall t \geq  0  \nonumber 
\end{eqnarray}
where $ c_{gas}$ is the O$_2$ concentration in the gas phase and $f_v$ is the ratio 
between the accessible 
free volume for O$_2$ in the silica network and the unit volume of the solid \cite{rigo}.
Initially there is a pure, c-Si substrate from $x=0$ to 
$\infty$, whose surface is exposed to a gaseous medium, O$_2$, with a 
relative concentration at this surface represented in the second  condition by 
$\frac{c_{gas}f_v}{c_{Si}^{bulk}}$. Reaction of these two species
 produces silicon oxide, not necessarily stoichiometric, as the solid-phase reaction Si + O$_2 \rightarrow$ SiO$_2$ may occur in several steps. As time proceeds,
O$_2$ diffuses through the oxide  network, reaches Si,  
reacts producing more oxide, and consequently pushes  the oxide/Si 
interface further in the positive $x$ region. 
Different values of $D$, $k$ and $c_{gas}$ imply 
different kinetics. However, the values of $D$ and $k$ may be used 
to define ``natural'' units for the system, resulting in  a set of 
 adimensional equations with only one parameter, namely $c_{gas}$ \cite{bark}. 
Under the following transformations:
\begin{eqnarray}
\label{natu}
\tau &=& t  k \nonumber \\
u &=& x \sqrt{\frac{k}{D}} \\
\phi_i(u,\tau)&=&\rho_i(x,t) \;\;\;\;\;\;\;\; \mbox{for } i=Si, O_2, oxide \nonumber
\end{eqnarray}
the resulting adimensional set of differential equations  and initial and boundary conditions are
\begin{eqnarray}
\label{eadi}
\frac{\partial{\phi_{O_2}}}{\partial{\tau}} &=& \frac{\partial^2{\phi_{O_2}}}{\partial u ^2} -  \phi_{O_2} \phi_{Si} \nonumber \\
\frac{\partial{\phi_{Si}}}{\partial{\tau}} &=&  -   \phi_{O_2} \phi_{Si}      \\
\phi_{Si}(u,0) &=& 1       \;\;\;\;\;      \forall u \geq 0  \nonumber \\
\phi_{O_2} (0,\tau) &=& \frac{c_{gas}f_v}{c_{Si}^{bulk}}    \;\;\;\;\  \forall \tau \geq  0  \nonumber
\end{eqnarray}
leaving $c_{gas}$ as the only parameter.
This is a powerful statement. 
It predicts that in all experiments performed at the same gas 
pressure, oxide growth kinetics and O$_2$ and Si concentrations versus depth curves (profiles) 
at different temperatures collapse to the same curves, provided 
that appropriate units are 
used: $1/k$ for time and $\sqrt{D/k}$ for thickness. 

In order to obtain profiles and growth kinetics,  eqs.(\ref{eadi}) 
must be solved. 
Those are partial, nonlinear, coupled differential equations and 
cannot be analytically solved at all times. 
On the other hand, they can be easily solved by finite 
differences methods, iterating for as long as it takes  to 
reproduce the
growth time of interest. The calculation gives then the profiles 
$\phi_i(u,\tau)$ of all species and the kinetics. Typical profiles at different growth times for a given value of $c_{gas}$ are shown 
in Figure 1. 
The O$_2$ profiles are not straight lines as they are assumed to be in
the steady-state regime \cite{deal}. 
Strictly speaking, 
steady-state is never attained as long as the oxide/Si interface 
moves deeper into the c-Si substrate. Furthermore, this interface 
is not sharp, in accordance with a wealth of experimental evidences 
\cite{feld2,guse,guse2}. 
Still in strict accordance with theoretical predictions 
\cite{pasq,mott,ston} and experimental observations 
\cite{himp,feld2,guse,guse2,sted}, a finite width region formed by 
Si in various lower oxidation states (+1, +2, +3, sub-oxides or Si-excess) 
early develops between the stoichiometric SiO$_2$ film and the c-Si 
substrate. In this way, O$_2$-Si reaction 
can take place in the whole region where non-fully oxidised 
Si is available. 

\begin{figure}
\includegraphics[width=8.cm, clip=true]{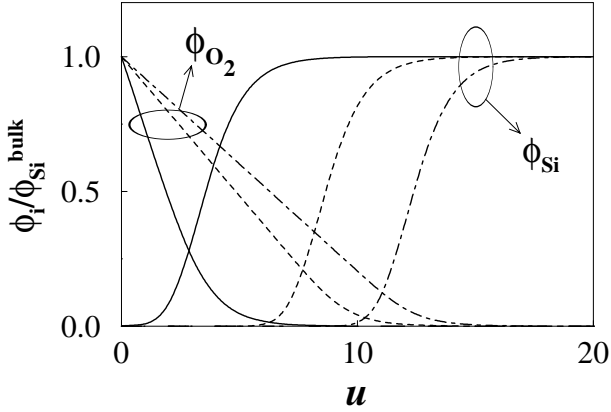}
\narrowtext
\caption{Calculated Si and O$_2$ profiles in the solid phase at
various temperatures and one pressure, for different oxidation times. 
The solid, dashed, and dot-dashed lines represent, respectively, 
the profiles at increasing oxidation times.}
\end{figure}

Fig. 1 shows that the two hypotheses of the Deal and Grove model 
are not  valid: i) Steady-state 
regime is never attained; ii) reaction between O$_2$ and Si 
leading to growth does not take place at a sharp oxide/Si 
interface. For high enough temperatures and pressures, and long enough 
times, 
the O$_2$ profile approaches a straight line, and the thickness of the 
near-interface, sub-oxide region is much smaller than the oxide thickness. This
is the observed situation at oxide thicknesses well above 20 nm.
In this limit, steady state and abrupt interface are plausible approximations.
In other words, 
asymptotically the Deal and Grove assumptions are valid, whereas 
at  initial stages they are not. The kinetics in the initial growth 
regime, which was described as ``anomalous'' in the
Deal and  Grove framework, naturally emerges from the present model as follows.
Let us consider the quantity that is mainly focused in 
practical situations: the oxide thickness $\chi$. We calculate it 
through the relation:
\begin{equation}
\label{xtau}
\chi(\tau) = \int_{0}^{\infty} \phi_{oxide}(u,\tau) du = \int_{0}^{\infty}(1-\phi_{Si}(u,\tau)) du
\end{equation}
The oxide growth kinetics $\chi(\tau)$ is completely determined
 by $\phi_{Si}(u,\tau)$, whose evolution is ruled by eqs.(\ref{eadi}) 
and therefore is determined by a unique parameter $c_{gas}$, the 
concentration of O$_2$ in the gas phase. However, due to the particular form of  the differential 
equations, together with boundary and initial conditions, the system 
presents a further scaling property. Suppose two different growth kinetics, 
with O$_2$ concentrations in the gas phase given by $c_{gas}^{(1)}$ 
and $c_{gas}^{(2)}=\alpha c_{gas}^{(1)}$, then  the following 
relations apply
\begin{eqnarray}
\label{scal}
\phi^{(2)}_ {O_2}(u,\tau)&=&\alpha \phi^{(1)}_ {O_2}(u, \alpha \tau) \nonumber \\
\phi^{(2)}_{Si}(u,\tau)&=& \phi^{(1)}_{Si}(u, \alpha \tau) \\
\chi^{(2)}(\tau) &= &\chi^{(1)}( \alpha \tau) \nonumber
\end{eqnarray}
Therefore, besides collapsing for different temperatures (i.e. different $D$ and $k$) 
when natural units of the system are used, kinetic curves for different $c_{gas}$ 
collapse to one single curve due to the symmetry present in the form of the diffusion-reaction 
model equations together with initial and boundary conditions. In summary,
the theoretical prediction is that 
all kinetic curves reduce to a single one. 

These predictions were tested with kinetic curves taken from 
references \cite{mass,sted,itoh,gane}, corresponding to oxidations in 
Joule-effect heated and rapid thermal processing furnaces in a variety of 
growth parameters (temperature, O$_2$  pressure, and time) leading to  oxide thicknesses ranging from 1 to 100 nm. In Fig. 2 
we show the collapse of the experimental
kinetics (symbols, colours) into the theoretical curve (line, black).
For the kinetics curves from reference \cite{itoh}, with five points in each 
curve, we first
used the scaling predicted by eqs.(\ref{scal}), obtaining four different 
curves, one for each temperature. We then scaled these curves using natural
units for $x$ and $t$ according to eqs.(\ref{natu}) until they overlapped
the theoretical curve. The  curves from reference  
\cite{mass} contained much more 
experimental points, 
extending from 0.1 to 1000 min, and from 1 to 80 nm. For these 
curves the pressure scaling predicted by
eqs.(\ref{scal}) is not perfect at long growth times, probably due to an increase in 
diffusivity. They were then collapsed one by one into the theoretical curve using eqs.(\ref{natu}) first, and then eqs.(\ref{scal}). The same procedure was used to collapse the kinetics from references \cite{sted,gane} into the theoretical curve.  
\begin{figure}
\includegraphics[width=9.cm, clip=true]{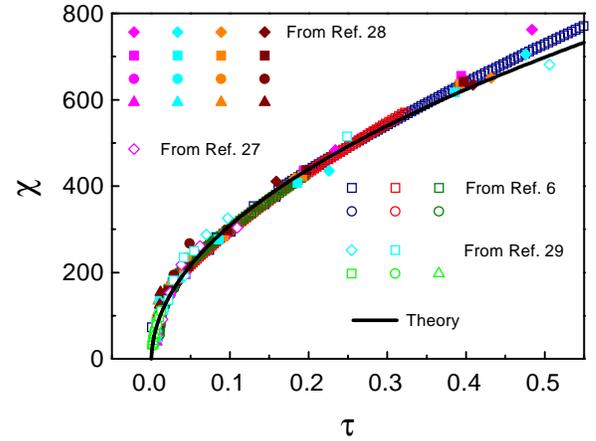}
\narrowtext
\caption{Calculated growth kinetics (solid line, black) and
experimental growth kinetics from the indicated references 
(symbols, colours) collapsed into the calculated curve.}
\end{figure}

The asymptotic depart from the theoretical curve is an expected result, since the
present model assumes a constant diffusivity throughout the whole growth
process, despite many evidences that this is not the case: i) 
transport of the diffusing species through the O-excess region near the 
surface \cite{trim,trim2,hama,stes}, through the Si-excess region 
near the oxide/Si interface \cite{lu,ston,guse,guse2}, and through the stoichiometric SiO$_2$ in the bulk of the growing oxide film 
have different
diffusivities; ii) the growth of the defective, near-surface and 
near-interface regions saturate within a few nm \cite{trim,himp,feld2,sted,stes}, while the bulk, 
stoichiometric oxide grows continuously. So, interstitial diffusion 
of O$_2$ through the bulk of the growing oxide gives a relative contribution that
increases as the width of this region becomes dominant; iii) oxidations 
performed in dry O$_2$ flow during long time intervals (hours) may suffer 
from water vapor contamination, which accelerates the oxide growth \cite{rigo,roch}.

Taking the dimensional 
oxide thickness  $X(t)=\chi(\tau) \sqrt{\frac{D}{k}}$, and defining  $\rho_O= \rho_{O_2} (0,t) = \frac{c_{gas}f_v}{c_{Si}^{bulk}}$,
the asymptotic solution to eqs.(\ref{eadi}) predicts the inverse growth 
rate $dt/dX =\frac{1.3}{D\rho_O}X$,
with $D $ constant, which is very similar to the asymptotic approximation
of the Deal and Grove model \cite{deal} 
$dt/dX =\frac{1}{D\rho_O}X$. In Fig. 3 the inverse growth rate is plotted 
as a function of oxide film thickness for 
different experimental kinetics \cite{mass}. 
\begin{figure}
\includegraphics[width=8.cm, clip=true]{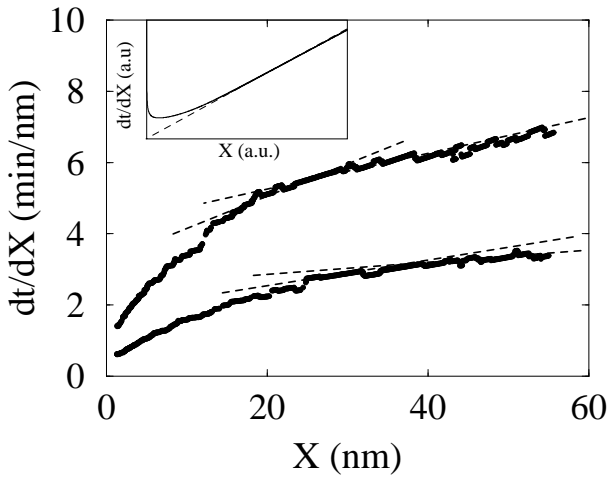}
\narrowtext
\caption{Experimental inverse growth rates [6] and theoretical inverse 
growth rate (inset).}
\end{figure}

A deviation from a straight line in the thick film range is apparent, corresponding to a variable, increasing
diffusivity, contrary to the theoretical prediction shown in the inset. 

Thus, the thermal growth of silicon oxide films on c-Si in dry O$_2$ is promoted by diffusion of O$_2$ through the growing oxide with variable diffusivity. Similar physical arguments could support a variable diffusion rate. Nevertheless, even using constant, effective 
values for the diffusivity and reaction rate,
modelling the growth as a diffusion-reaction process where conditions such as initial 
thickness, steady-state regime, and reaction solely at an abrupt interface 
are abandoned allows to reproduce the
observed kinetics, especially in the lower thickness (ultrathin film) regime
which is the one of present and future practical interest. Realistic model evaluations of 
variable diffusivity and reaction rate stay beyond the 
scope of the present work, although this seems to be the only route to 
a quantitative description of the complete growth kinetics.

The present approach opens the
possibility of incorporating
new and determinant 
facts on O$_2$ transport at initial stages of thermal oxide 
growth and on structure of the oxide/Si interface, that were discovered after the proposal of
Deal and Grove \cite{deal} and not yet integrated into growth kinetics 
models. Some were discussed here, like
the graded, Si-excess nature of the oxide/Si interface and the 
incorporation of freshly arriving O$_2$ into different regions of the 
growing oxide, while several others may also be included in a diffusion-reaction
approach, such as the role played by oxygen excess centers 
\cite{stes} in the near-surface region on oxygen transport
which has been recently investigated \cite{hama}.

\section*{Acknowledgements}
We thank M. A. Idiart for fruitful discussions and suggestions. 
This work has been partially supported by brazilian agencies CNPq, FINEP and FAPERGS.
\end{multicols}
\end{document}